\newif\ifpdflatex    
\pdflatextrue           

\documentclass[fleqn,usenatbib,useAMS,twocolumn]{aastex631}
\usepackage{amsmath,esint}
\usepackage{url,aasmacros}
\usepackage{natbib}

\usepackage{soul, CJK}
\usepackage{multirow}
\usepackage{tablefootnote}

\usepackage{bm}
\usepackage{xspace}
\usepackage{color}
\usepackage{enumitem}
\usepackage{booktabs}
\usepackage{ragged2e}

\usepackage{graphicx}	
\usepackage{amsmath}	
\usepackage{amssymb}	
\usepackage{bm}		
\usepackage{url}
\usepackage{amsbsy}
\usepackage{xspace}
\usepackage{hyperref}
\usepackage{refcount}
\usepackage{longtable}
\usepackage[colorinlistoftodos, textsize=scriptsize]{todonotes}
\usepackage[flushleft]{threeparttable}
\usepackage{longtable}

\usepackage[T1]{fontenc}
\usepackage{ae,aecompl}

\def\lesssim{\mathrel{\hbox{\rlap{\hbox{\lower5pt\hbox{$\sim$}}}\hbox{$<$}}}}
\def\gtrsim{\mathrel{\hbox{\rlap{\hbox{\lower5pt\hbox{$\sim$}}}\hbox{$>$}}}}

\def\Rsun{{\rm R$_{\odot}$}\xspace}
\def\Msun{{\rm M$_{\odot}$}\xspace}

\def\coa{\rm ${}^{12}$C${}^{16}$O\xspace}
\def\cob{\rm ${}^{12}$C${}^{18}$O\xspace}

\newcommand{\angstrom}{\textup{\AA}\xspace}

\usepackage{upgreek}                                                  
\usepackage{xspace}                                                       
\newcommand{\um}{$\upmu$m\xspace}            
\def\apjl{ApJ}%
\def\apjs{ApJS}%
\def\aap{A\&A}%
\shorttitle{Oxygen Isotope Ratios in HdC Stars}
\shortauthors{Mehla et al.}

\begin{document}
\title{Oxygen Isotope Ratios in Hydrogen-Deficient Carbon Stars: \\ A Correlation with Effective Temperature and Implications for White Dwarf Merger Outcomes}
\author[0000-0002-3155-6584]{Advait Mehla}
\affil{Department of Physics, Indian Institute of Technology Bombay, Powai 400076, India}
\author[0000-0002-5619-4938]{Mansi M. Kasliwal}
\affil{Division of Physics, Mathematics, and Astronomy, California Institute of Technology, Pasadena, CA 91125, USA}
\author[0000-0003-2758-159X]{Viraj Karambelkar}
\altaffiliation{Visiting Astronomer at the Infrared Telescope Facility, which is operated by the University of Hawaii under contract 80HQTR24DA010 with the National Aeronautics and Space Administration.}
\affil{Division of Physics, Mathematics, and Astronomy, California Institute of Technology, Pasadena, CA 91125, USA}
\author[0000-0003-4237-0520]{Patrick Tisserand}
\affil{Sorbonne Universit\'es, UPMC Univ. Paris 6 et CNRS, UMR 7095, Institut d'Astrophysique de Paris, IAP, 75014 Paris, France}
\author[0000-0002-7654-7438]{Courtney Crawford}
\affil{Sydney Institute for Astronomy (SIfA), School of Physics, University of Sydney, NSW 2006, Australia}
\author[0000-0002-0141-7436]{Geoffrey Clayton}
\affil{Space Science Institute, 4765 Walnut St, Suite B, Boulder, CO 80301, USA}
\author[0000-0001-9226-4043]{Jamie Soon}
\affil{Research School of Astronomy and Astrophysics, Australian National University, Cotter Rd, Weston Creek ACT 2611, Australia}
\author[0000-0002-6112-7609]{Varun Bhalerao}
\affil{Department of Physics, Indian Institute of Technology Bombay, Powai 400076, India}

\begin{abstract}
    
    Hydrogen-deficient Carbon (HdC) stars are a class of supergiants with anomalous chemical compositions, suggesting that they are remnants of CO-He white dwarf (WD) mergers. This class comprises two spectroscopically similar subclasses -- dusty R Coronae Borealis (RCB) and dustless Hydrogen-deficient Carbon (dLHdC) stars. Both subclasses have a stark overabundance of $^{18}$O in their atmospheres, but spectroscopic differences between them remain poorly studied. 
    We present high-resolution ($R \approx 75000$) \textit{K}-band spectra of six RCB and six dLHdC stars, including four newly discovered dLHdC stars, making this the largest sample to date. We develop a semi-automated fitting routine to measure $^{16}$O/$^{18}$O ratios for this sample, tripling the number of dLHdC stars with oxygen isotope ratios measured from high resolution spectra. All six dLHdC stars have $^{16}$O/$^{18}$O $< 1$, while the RCB stars have $^{16}$O/$^{18}$O~$>4$. Additionally, for the first time, we find a trend of decreasing $^{16}$O/$^{18}$O ratios with increasing effective temperature for HdC stars, consistent with predictions of theoretical WD merger models. 
    However, we note that current models overpredict the low $^{16}$O/$^{18}$O ratios of dLHdC stars by two orders of magnitude. 
    We also measure abundances of C, N, O, Fe, S, Si, Mg, Na, and Ca for these stars.
    We observe a correlation between the abundances of $^{14}$N and $^{18}$O in our sample, suggesting that a fixed fraction of the $^{14}$N is converted to $^{18}$O in these stars via $\alpha$-capture. Our results affirm the emerging picture that the mass ratio/total mass of the WD binary determine whether an RCB or dLHdC is formed post-merger.
\end{abstract}

\section{Introduction}
\label{sec:intro}

Hydrogen-deficient Carbon (HdC) stars are a class of supergiant stars with unusual chemical compositions,  characterized by an acute deficiency of hydrogen and an overabundance of carbon \citep{clayton1996, 2012JAVSO..40..539C, Lambert1994}. Their peculiar chemical compositions suggest that they originate in the merger of a CO-core and an He-core white dwarf \citep{clayton2007} in what is known as the double-degenerate (DD) formation scenario. They are found in all old galactic sub-structures, the thick disk, bulge and halo, but also in the younger thin disk \citep{tisserand2024}. Adding to their intrigue, HdC stars seem to comprise two varieties with similar chemical compositions but distinct photometric properties: R Coronae Borealis (RCB)-type stars that show erratic brightness variations resulting from dust {obscuration} episodes \citep{tiss2020}, and dustless Hydrogen-deficient Carbon (dLHdC) stars that do not show dramatic brightness variations or any signs of dust formation \citep{tisserand2022}. Why RCB stars form dust while dLHdC stars do not despite having similar chemical compositions is still a mystery.

Several studies have attempted to identify chemical differences between RCBs and dLHdC stars as a step towards solving their dusty mystery. Using medium resolution K-band spectra of these stars covering the CO molecular absorption bandheads, \citet{clayton2007} first noted that RCB stars and dLHdC stars possibly differ in their oxygen isotope ratios, with dLHdCs having $^{16}$O/$^{18}$O $< 1$ {and} RCBs having $^{16}$O/$^{18}$O $>1$. This was confirmed with high-resolution spectra by \citet{Garcia-Hernandez2009, Garcia-Hernandez2010}. However, these studies were based on small samples of five RCB stars and two dLHdC stars. Although there were over fifty RCB stars known at the time, only four dLHdC stars were known, of which two were too warm to exhibit CO lines. However, this has changed recently with the discovery of 27 new dLHdC stars \citep{tisserand2022}. Additionally, a large number of RCB stars have been observed at near-infrared wavelengths \citep{karambelkar2021, Karambelkar2024}, enabling spectroscopic comparisons of larger samples of RCB and dLHdC stars. 

Using medium resolution NIR spectra for the expanded sample of RCBs and dLHdCs, \citet{karambelkar2022} showed that dLHdC stars generally tend to have lower values of $^{16}$O/$^{18}$O than RCBs, and suggested that this difference could be attributed to different masses or mass-ratios of the merging white-dwarfs \citep{Crawford2024}. However, the oxygen isotope ratios measured from the medium resolution spectra have large uncertainties due to excessive blending with molecular CN lines in this wavelength range. This lack of precision precludes any further insights, such as quantitative comparisons with theoretical models or identification of trends within the data. 

In this paper, we present high resolution (R$\approx$75000) K-band spectra of six RCB and six dLHdC stars. These are the highest resolution NIR spectra of the largest sample of RCB and dLHdC stars to date. We use these spectra to derive the oxygen isotope ratios and other elemental abundances of these stars and examine differences between RCB and dLHdC stars. The paper is structured as follows -- Section \ref{sec:observations} describes the spectroscopic observations, Section \ref{sec:analysis} describes the abundance measurement method, {and Section} \ref{sec:discussion} examines the differences and trends of the abundance measurements. We conclude with a summary of our results in Section \ref{sec:conclusion}.

\begin{table*}[t]
    \centering
    \begin{threeparttable}
        \caption{Log of spectroscopic observations.} 
        \label{obslog}
        \begin{tabular}{lcccccc}
            \hline \\[-3ex]
            \hline
            Name                 & Class & Date (UT)  & S/N & Telluric Standard & Total Exposure Time (min) &  Number of Coadds     \\ \hline
            HD 137613            & dLHdC & 2022-09-14 & 104 & HIP79881          & 5.0                 &       5                \\
            A223*                & dLHdC & 2022-09-14 & 127 & HIP93667          & 60.0                &       6                \\
            B42*                 & dLHdC & 2022-09-14 &  32 & HIP93667          & 20.0                &       2               \\
            HD 182040            & dLHdC & 2022-09-22 & 108 & HIP95793          & 3.0                 &        6               \\
            C38*                 & dLHdC & 2022-09-23 &  73 & HIP93667          & 100.0               &       10                 \\
            B566*                & dLHdC & 2022-09-30 &  78 & HIP91137          & 90.0                &       9                \\
            WISE J1820+         & RCB   & 2022-09-24 &  41 & HIP93667           & 24.0                &       12                \\
            WISE J1942+          & RCB   & 2022-09-22 &  98 & HIP98953          & 50.0                &       5                \\
            IRAS 1813+           & RCB   & 2022-09-22 & 149 & HIP95793          & 50.0                &       5                \\
            ASAS-RCB-18          & RCB   & 2022-09-23 &  79 & HIP95793          & 13.5                &       9                \\
            V CrA                & RCB   & 2022-09-23 & 136 & HIP93470          & 10.0                &        5               \\
            AO Her               & RCB   & 2022-09-24 &  88 & HIP91315          & 10.0                &      10                 \\
            ASAS-RCB-21          & RCB   & 2022-09-24 &  81 & HIP95793          & 28.0                &      14                 \\
            NSV 11154            & RCB   & 2022-09-24 &  87 & HIP91315          & 54.0                &      5                 \\
            WISE J1818+         & RCB   & 2022-09-30 &  68 & HIP93667           & 7.5                 &      5                 \\ \hline
        \end{tabular}
        \begin{tablenotes}
            \footnotesize
            \item[] \textbf{Note:} The recently discovered dLHdC stars \citep{tisserand2022} are marked by asterisks. \\
            The following abbreviations are used throughout the text:
            WISE J194218.38-203247.5: WISE J1942+; IRAS 1813.5-2419: IRAS 1813+; WISE J181836.38-181732.8: WISE J1818+; WISE J182010.96-193453.4: WISE J1820+. The last two stars are referred to as WISE-ToI-222 and WISE-ToI-223 in {\citet{karambelkar2021} and \citet{tiss2020}}.
        \end{tablenotes}
    \end{threeparttable}
\end{table*}
\setcounter{table}{2}
\begin{table*}[t]
    \centering
    \begin{threeparttable}
        \caption{Adopted stellar parameters.} 
        \label{stellarparams}
        \begin{tabular}{lcccc}
            \hline \\[-3ex]
            \hline
            Name         & T$_\textrm{eff}$ (K)\tnote{a} & $\log g$ (cm s$^{-2}$) & $\xi$ (km s$^{-1}$) & $v_{\text{mac}}$ (km s$^{-1}$) \\ \hline
            HD 137613    & 5500                &          1.0        &       6.5           &  0.0  \\
            HD 182040    & 5750                          &          1.0        &       6.0           &  0.0  \\
            A223         & 6250                          &          1.0        &       7.0           &  0.0  \\
            B566         & 5750                          &          1.0        &       6.0           &  0.0  \\
            B42          & 5500                          &          1.0        &       6.0           &  0.0  \\
            C38          & 5750                          &          1.0        &       6.0           &  0.0  \\
            AO Her       & 4750                 &          1.0        &       7.0           &  8.0 \\
            ASAS-RCB-18  & 5000                          &          1.0        &       7.0           &  6.0 \\
            ASAS-RCB-21  & 5000                 &          1.0        &       7.0           &  8.0 \\
            NSV 11154    & 5250                          &          1.0        &       6.0           &  8.0 \\
            WISE J1818+ & 5000                          &          1.0        &       7.0           &  8.0 \\
            WISE J1942+  & 4500                 &          1.0        &       7.0           &  6.0 \\
        \end{tabular}
        \begin{tablenotes}
            \footnotesize
            \item[a] Chosen values are the nearest available from our grid of models. Values are adopted using a combination of empirical color-temperature calibration \citep{Crawford2023} and estimates from spectral energy distribution fits from literature \citep{bergeat2001, tisserand2012, karambelkar2021} where available.
        \end{tablenotes}
    \end{threeparttable}
\end{table*}

\section{Observations}
\label{sec:observations}
Our data consists of high-resolution ($R \approx 75000$ or 4 km {s$^{-1}$}) \hbox{\textit{K}-band} spectra of nine RCB and six dLHdC stars acquired with the iSHELL spectrograph \citep{ishell} on the 3.2 m NASA Infrared Telescope Facility (IRTF) at Mauna Kea Observatory, Hawaii. The observations were obtained {under program 2022B065 (PI: Karambelkar)} over six nights between 2022 September 14 and 2022 October 3, with the 0.375" slit, using the \textit{K} order sorter in the \textit{K}$_3$ mode which covers the wavelength range of $2.25 - 2.48$~\um. This region was chosen to include multiple bandheads of $^{12}$C$^{16}$O and ${}^{12}$C${}^{18}$O, as well as several lines of ${}^{12}$C${}^{14}$N. The full log of observations is listed in Table \ref{obslog}{, along with the median continuum signal-to-noise ratios in the 2.3 -- 2.4 \um region}. {Spectra were reduced} using the \texttt{idl} software \texttt{spextool} \citep{spextool}, and were calibrated and corrected for telluric absorption using the \texttt{xtellcor} package \citep{xtellcor}. 
{We used the in-built spextool module \texttt{xspextool} for dark and flat calibration, wavelength calibration using arc-lamp observations and extraction of the 1D spectrum in each science exposure. Multiple exposures of the same target were combined using the \texttt{xcombspec} submodule. Once telluric corrected, the different echelle orders were merged using the submodule \texttt{xmergeorders}, that scales the orders to match flux levels of overlapping wavelengths in successive orders. Continuum-normalized spectra were obtained using a custom \texttt{python} algorithm that identifies the continuum level by identifying regions relatively free of strong molecular absorption and fits a fourth-order polynomial to them. The spectra were doppler corrected to the rest frame of each star using radial velocity measurements and manually correcting offsets of up to 20--30 km s$^{-1}$ caused by pulsations that are common in these stars \citep{tisserand2024}.}

For the RCB stars in our sample, we determined photometric phases at the epochs of the IRTF observations using time-series photometric information aggregated from various surveys by the DREAMS monitoring portal\footnote{\url{https://dreams.anu.edu.au/monitoring/}}. Of the nine RCB stars, IRAS 1813+ and WISE J1820+ were observed in a decline, where their V-band brightness were fainter by $\Delta V \approx 1$ and $\Delta V \approx 3$ mag than maximum luminosity respectively. Subsequently, the spectra for these two stars have not been analyzed as the photospheric lines are heavily diluted {due to a decrease of the photospheric to circumstellar dust shell flux ratio in the \textit{K}-band} caused by newly ejected dust clouds. Additionally, although it was observed at maximum light, V CrA was found to have a featureless spectrum, likely due to a combination of its high estimated effective temperature ($\sim 6500$ K) and large \textit{K}-band excess from circumstellar dust.
ASAS-RCB-18 and ASAS-RCB-21 were recovering from decline phases and were both roughly 0.8 mag fainter than maximum luminosity, but this does not appear to appreciably dilute the photospheric spectrum, as we see lines that are stronger than or comparable to other stars {despite a similar V-band {deficit} as IRAS 1813+. To understand this apparent discrepancy, we examined mid-infrared NEOWISE W1 and W2 light curves \citep{mainzer2014}, that track variations of emission from the circumstellar dust shell \citep{feast1997}. The NEOWISE data shows that IRAS\,1813+ was at its maximum mid-IR brightness at the time of the iSHELL observation, while ASAS-RCB-18 and 21 were at minimum mid-IR brightnesses, suggesting that the extent of dilution observed in the \textit{K}-band is substantially lower for ASAS-RCB-18 and 21. For this reason, we include these objects in our analysis.}
The remaining four RCB stars were observed at maximum light, bringing the total number of RCB stars analyzed in this work to six.

Among the dLHdC stars, \citet{tisserand2024} identified four as part of the galactic bulge structure (A223, B42, C38, and B566), one as belonging to the thick disk (HD 137613), and another (HD 182040) likely associated with the thin disk. For the RCB stars, only one has been assigned to a galactic substructure: NSV 11154, classified as a halo star due to its high galactic latitude and significant radial velocity.
\begin{figure*}
    \centering
        \includegraphics[width=0.4\textwidth]{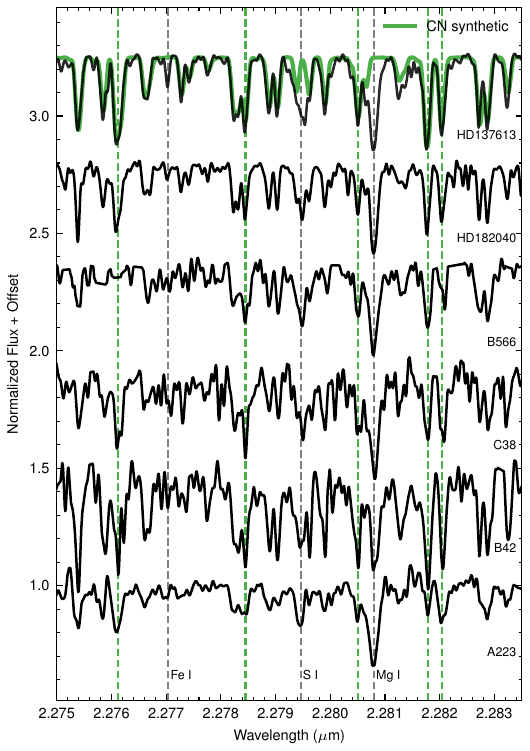}
        \includegraphics[width=0.4\textwidth]{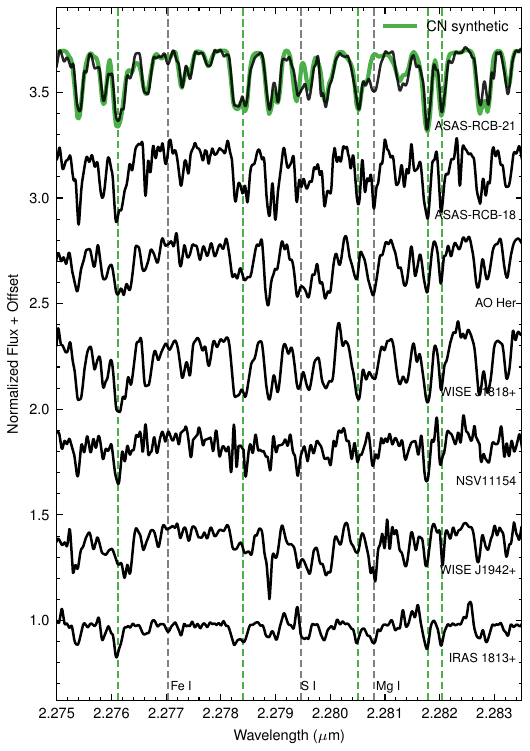}
        \caption{Spectra centred at $\sim$2.28 \um showing a region with $^{12}$C$^{14}$N lines for the dLHdC stars (left) and RCB stars (right). $^{12}$C$^{14}$N synthetic spectra generated for HD 137613 and ASAS-RCB-21 are overplotted in green for comparison, with a few strong lines marked. Also marked in gray are the locations of a few atomic lines.} \label{fig:CN}
        \includegraphics[width=0.4\textwidth]{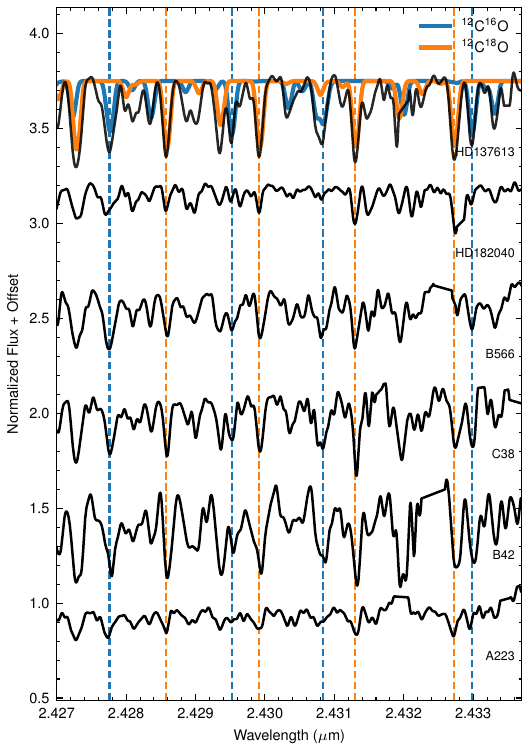}
        \includegraphics[width=0.4\textwidth]{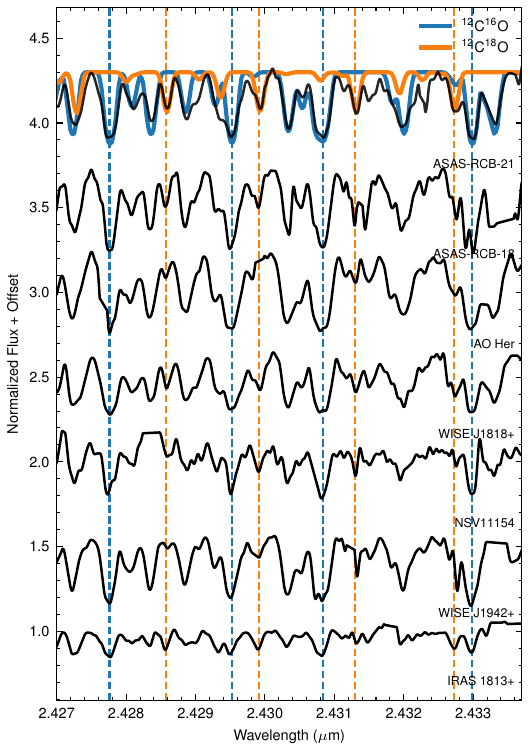}
        \caption{Spectra centred at $\sim$2.43 \um showing a region dominated by $^{12}$C$^{16}$O and $^{12}$C$^{18}$O lines for the dLHdC stars (left) and RCB stars (right). Individual synthetic spectra generated for HD 137613 and ASAS-RCB-21 are overplotted in blue and orange for comparison, with a few strong lines marked. It is clear that $^{12}$C$^{18}$O is the dominant isotopologue in all dLHdCs, while $^{12}$C$^{16}$O is much more prominent in RCBs.} \label{fig:CO}
    \end{figure*}

\section{Abundance Analysis}  \label{sec:analysis}
\subsection{Notable Spectral Features}
\setcounter{table}{1}
\begin{table}[t]
    \centering
        
        \begin{tabular}{lc}
            \hline \\[-3ex]
            \hline
            Species & Wavelengths (\um) \\ \hline
            S~I     & 2.2655, 2.2707, 2.2875 \\ 
            Si~I    & 2.2665, 2.275, 2.3142, 2.3953, 2.4245, 2.4575 \\
            Fe~I    & 2.262, 2.277, 2.2835, 2.331, 2.4164, 2.4265 \\
            Na~I    & 2.3348, 2.338  \\
            Mg~I    & 2.2808, 2.2906 \\
            C~I     & 2.2906, 2.3091, 2.3154, 2.3442 \\
            ${}^{12}$C${}^{16}$O & 2.293, 2.322, 2.352, 2.382, 2.414, 2.445, 2.478 \\
            ${}^{12}$C${}^{18}$O & 2.349, 2.378, 2.408, 2.438, 2.469 \\
            \hline
            
        \end{tabular}
        \caption{Locations of prominent features in our spectra. {See Table \ref{tab:clist} for the full C~I line list.}}
        \label{atomiclines}
\end{table}
\setcounter{table}{3}

Our spectra cover several molecular absorption features such as the CN Red System of ${}^{12}$C${}^{14}$N, the first seven overtone bandheads of ${}^{12}$C${}^{16}$O, the first five overtones of ${}^{12}$C${}^{18}$O (listed in Table \ref{atomiclines}), and C$_2$ absorption lines from the Phillips and Ballik-Ramsay systems. There is considerable blending of lines especially beyond 2.29 \um. Figure \ref{fig:CN} shows a region with prominent CN lines for both dLHdC and RCB stars, along with a CN synthetic spectrum generated for one star in each class. Figure \ref{fig:CO} shows a region dominated by ${}^{12}$C${}^{16}$O and ${}^{12}$C${}^{18}$O lines for the dLHdC and RCB stars. We also have several atomic lines in our spectra, however most of them are severely blended with molecular features. A few prominent ones that we use for determining elemental abundances are listed in Table \ref{atomiclines}. {C~I lines are only detected in the dLHdC stars in our sample.}

Our wavelength range also includes bandheads of ${}^{13}$C${}^{16}$O, but these are not detected in any star in our sample. This indicates a high $^{12}$C/$^{13}$C ratio that is consistent with the DD scenario and other observations of HdC stars \citep{clayton2007,Garcia-Hernandez2009,Jeffery2011}. We also searched for 1 -- 0 lines of HF (using ExoMol linelist by \citealt{TENNYSON2024109083}) which were recently detected in the \textit{K}-band spectrum of DY Per \citep{agh2023} but found none. This is consistent with \citet{agh2023} observing a lack of these lines in the RCB WX~CrA and other HdC stars. They note that the absence of HF lines is likely indicative of a severe H-deficiency rather than a F-deficiency in these stars, as the abundances of both affect the formation of HF in the atmosphere.

\subsection{Possible CO Emission Line Contamination} \label{sec:emission}
\begin{figure*}[t]
    \centering
    \includegraphics[width = \textwidth]{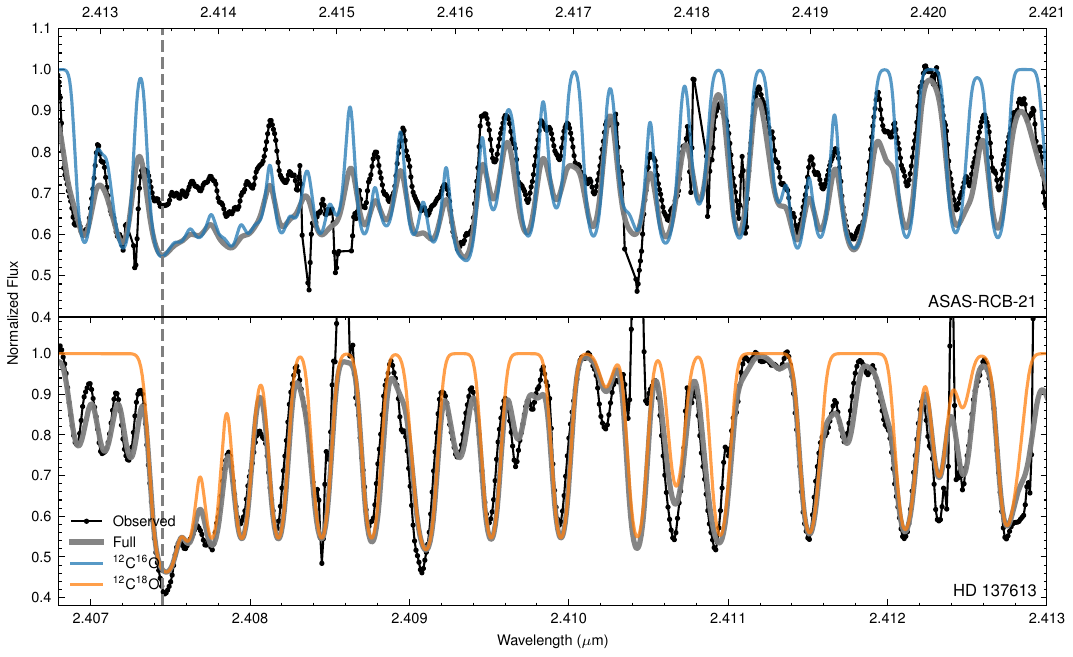}
    \caption{A comparison of strong bandhead of \coa (ASAS-RCB-21, top) and \cob (HD 137613, bottom) highlighting the distortions of stronger lines observed in RCB stars, likely due to contamination from circumstellar emission lines (see Section \ref{sec:emission}). The location of the respective bandheads is indicated by a dashed vertical line. These regions also have several artifacts (near 2.415, 2.4175, 2.418 \um in the top; 2.4085, 2.4105, 2.4125 \um in the bottom) related to erroneous telluric correction.} \label{fig:bandhead}
\end{figure*}

We note that for the RCB stars in our sample, strong ${}^{12}$C${}^{16}$O absorption lines beyond 2.35 \um appear flattened or distorted. While the wings of these lines match the profiles of model synthetic spectra, the cores appear truncated. This is seen in all RCBs in our sample (most prominently in ASAS-RCB-21) but in none of the dLHdCs. This truncation is most noticeable at strong ${}^{12}$C${}^{16}$O absorption bandheads and gets more pronounced towards redder wavelengths. Figure \ref{fig:bandhead} compares prominent ${}^{12}$C${}^{16}$O bandheads in ASAS-RCB-21 and HD~137613, illustrating clear distortions from the synthetic spectrum in the RCB star, while the dLHdC shows a near-perfect match with the synthetic spectrum.

Dilution of absorption bands due to circumstellar dust is unlikely to explain this as nearby absorption lines of similar strength around the truncated CO bandhead are unaffected. We expect dilution to uniformly weaken the spectra of RCB stars, and this is discussed in Section \ref{sec:dilution}. 
Instead, we suggest that this truncation is due to the absorption lines being `filled out' by CO emission lines from circumstellar molecular gas around the RCB stars. Narrow emission from circumstellar C$_{2}$ and CN molecules has been previously detected in RCB stars \citep{rao1999, rao2004} during the decline phase. This circumstellar emission is believed to be persistent \citep{rao2004}, and can contaminate photospheric absorption lines even if the star is observed at maximum light. 
We expect the strongest lines to be the most affected, as their corresponding circumstellar emission lines would also be strong -- this explains the severity at bandheads. Conversely, weak lines are unaffected. We have had some success in replicating such line profiles with a toy model that combines CO emission lines obtained by inverting an absorption spectrum, confirming that this is a plausible explanation.

It is important to emphasize that this does not seem to affect the entire spectrum. We note that lines of the 2~--~0 and 3~--~1 bandheads are unaffected, while the subsequent bandheads ($\lambda > 2.35$ \um) show progressively more severe effects at redder wavelengths. This could be related to how energy levels are populated by collisional excitation which likely causes these emission lines \citep{rao1999}, and is an interesting avenue for future studies. We exclude such lines from our analysis and they do not affect our abundance measurements.

\subsection{Spectral Modeling}
We constructed model synthetic spectra using a grid of hydrogen-deficient, spherically symmetric MARCS (Model Atmospheres in Radiative and Convective Scheme) atmospheric models, generated with input abundances of $\log \epsilon(\text{H}) = 7.5$, $\log \epsilon(\text{He}) = 11.5$, $\log \epsilon(\text{C}) = 9.5$ (C/He $= 0.01$), and $\log \epsilon(\text{O}) = 8.8$ \citep{gust1975,gust2008,bell1976,plez2008}. The models were constructed for five different nitrogen abundance values of $\log \epsilon(\text{N}) = 7.0,\ 7.5,\ 8.0,\ 8.5$, and $9.4$\footnote{Note: $\log \epsilon(\text{X})$ refers to abundances normalized to \hbox{$\log \sum_{i}^{} \mu_i \epsilon_i = 12.15$}, where $\mu_i$ is the mean atomic weight of an element i.  $\log \epsilon(\text{X})$ will be denoted by $A(X)$ throughout the text. }. A solar metallicity is assumed for other elements. The models assume a stellar mass of $1 \ M_{\odot}$, surface gravity values ($\log g$) $= 1.0$, microturbulence \hbox{($\xi$) $= 5$ km s$^{-1}$}, and were computed for a grid of effective temperatures (T$_\textrm{eff}$) ranging from $4000 \text{ -- } 7500$ K (in steps of $250$ K). 

We generated synthetic spectra using \texttt{TSFitPy} \citep{tsfit} -- a \texttt{python} wrapper around the radiative transfer package \texttt{Turbospectrum~v20.0} \citep{ts0, ts1, ts20}. We used the CO line list provided by B. Plez (priv. commun. and described in \citealt{1994ApJS...95..535G}) and the ExoMol linelists for CN and C$_2$ \citep{2018MNRAS.480.3397Y,TENNYSON2024109083}. Atomic line lists are generated using the VALD database \citep{vald}. {We found several irregularities in the C~I line lists available in the literature (as noted previously by \citealt{Garcia-Hernandez2009}), and used a subset of lines from the VALD-Kurucz database that are present in our data. These are reported in Table \ref{tab:clist}.}

{While \texttt{Turbospectrum v20.0} includes support for Non-Local Thermodynamic Equilibrium (NLTE) spectrum synthesis, the requisite departure coefficients and atomic models are only available for a limited set of atoms (Ca, Fe, Mg, Na, O, Si at the time of writing). NLTE calculations for molecules are not supported at this time, and given that the main focus of our work is to recover abundances from molecular lines, we only use Local Thermodynamic Equilibrium (LTE) spectrum synthesis in our analysis. }

\subsection{Spectral Fitting} \label{sec:fitting}

\begin{figure*}
    \centering
    \includegraphics[width = \textwidth]{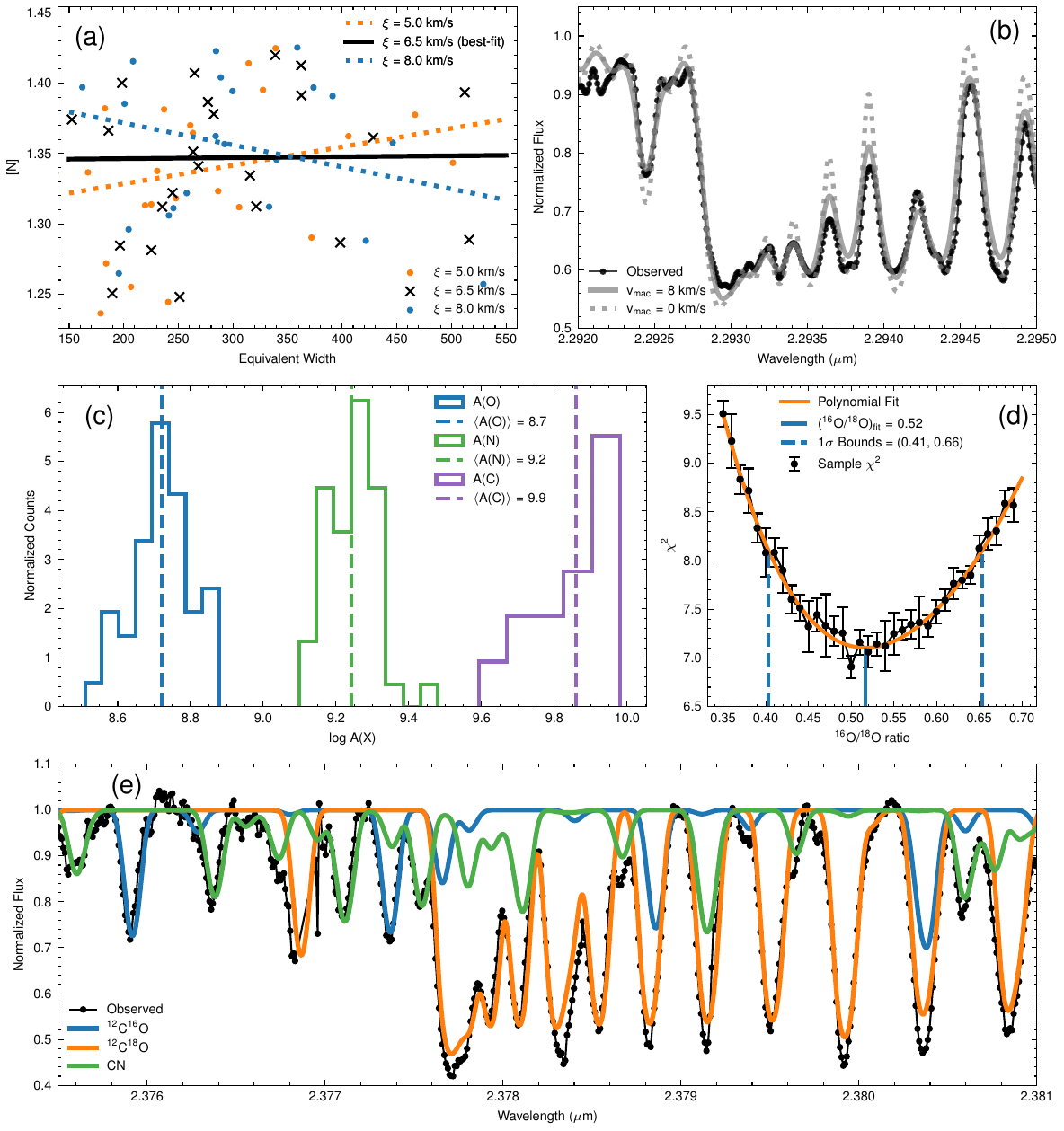}
    \caption{
    These panels show the steps discussed in Section \ref{sec:fitting} for HD 137613, except (b) which shows WISE J1818+ to illustrate {the additional broadening due to macroturbulence observed in RCBs}. \\
    (a) Equivalent Width v/s [N] obtained for 3 different $\xi$ values, color-coded as per the legend with linear fits made to visually illustrate that the quantities are uncorrelated at the best-fit $\xi = 6.5$ km/s. \\
    (b) The spectrum of RCB WISE J1818+ near the 2-0 \coa bandhead compared with synthetic spectra generated for a $v_{mac}$ of 0 and 8 km/s, indicating the presence of additional broadening. \\
    (c) Distributions of fits obtained for $A(O)$ and $A(N)$ from the respective linemasks, with mean values marked. \\
    (d) $\chi^2$ curve obtained by fitting a range of $^{16}$O/$^{18}$O ratios, with the best-fit value and $1\sigma$ bounds marked. The $\chi^2$ values have errorbars as the final fits have an inherent spread caused by random sampling of initial guesses.\\
    (e) Synthetic spectra generated for individual molecules compared with the observed data, indicating they are in agreement. \\   }
    \label{fig:fitting}
    \end{figure*}

We fit the models to our observed spectra to derive the abundances of different elements. In addition to the elemental abundances, the model fitting requires four additional stellar parameters -- effective temperature (T$_\textrm{eff}$), surface gravity ($\log g$), microturbulence ($\xi$), and macroturbulence ($v_{\text{mac}}$). The micro- and macroturbulence velocities are free parameters in our fitting and are estimated using the observed spectra (see below for details). We fix $\log g = 1.0$ for all stars, computed by assuming $M = 0.8$~\Msun, T$_\textrm{eff} = 5000$~K, $M(V)=-3$ and $R=47$~\Rsun. We use a combination of an empirical color-temperature calibration \citep{Crawford2023} along with estimates from spectral energy distribution fits in literature where available to estimate T$_\textrm{eff}$ {-- this is another fixed parameter that cannot be determined from our spectral fits}. Table \ref{stellarparams} lists the stellar parameters adopted for the stars analyzed here. We discuss how potential uncertainties in these values might propagate to derived abundances in Section \ref{sec:errors}.  

We convolve the synthetic spectra with a Gaussian function to match the iSHELL spectral resolution and fit the convolved models to our observed spectra using $\chi^2$ minimization implemented in \texttt{TSFitPy}. We made some modifications to the default version of \texttt{TSFitPy} to enable it to generate synthetic spectra from our custom, sparse model grid. Specifically, we suppressed the interpolation of metallicity, $\log g$, and T$_\textrm{eff}$. We also modified the code to use $^{16}$O/$^{18}$O ratio as a free parameter for fitting.

The carbon, nitrogen, oxygen abundances and oxygen isotope ratios for each star in our sample are derived using the following steps (illustrated in Figure \ref{fig:fitting} for HD~137613) -- 
\begin{enumerate}
    \item An inital guess for the different elemental abundances is obtained by visually comparing the models with the observed lines. 
    \item The microturbulence ($\xi$) is determined using the method outined in \citet{1984A&A...134..189M}. A set of single, unblended CN lines (or CO lines if a star has low $A(N)$) is identified. Models are fit to these lines to measure N (or O) abundances assuming different values of $\xi$. The value of $\xi$ for which the correlation between derived abundances and the line equivalent widths is closest to zero is adopted as the microturbulence value for the star (see Figure~\ref{fig:fitting}a).
    \item Clusters of partially blended lines are used to determine $v_{\text{mac}}$, as the profiles of such regions show the most prominent effects of any additional broadening. We find that dLHdC stars do not show any broadening beyond instrumental resolution, while RCB stars indicate \hbox{$v_{\text{mac}}\sim$  6 -- 8  km s$^{-1}$}. This is illustrated in Figure~\ref{fig:fitting}b for the RCB star WISE J1818+.
    \item Next, $A(C)$ is measured using 3 -- 4 unblended C$_2$ lines that are available in our spectra. Since the vast majority of C$_2$ lines happen to be blended with CN, this serves as an initial estimate.  \\
    \item The obtained $A(C)$ value is then used to fit a set of 30 -- 40 CN lines to measure the nitrogen abundance. This is then fed back to a larger C$_2$ linemask which includes blended lines to further refine the C abundance. If required, this process is iterated a few times until the derived values converge. The uncertainty on the C and N abundances is obtained by measuring the standard deviation of the line-to-line scatter in the measured abundances. {{Figure~\ref{fig:fitting}c shows the distributions of fitted abundances obtained for C, N, and O.}} \\
    \item Finally, the oxygen abundance and isotope ratios are measured using a set of ${}^{12}$C${}^{16}$O and ${}^{12}$C${}^{18}$O lines. To measure the oxygen-isotope ratios, we first measure the goodness-of-fit ($\chi^2$) by fitting synthetic spectra to all selected absorption lines simultaneously for a range of input $^{16}$O/$^{18}$O values, with the total oxygen abundance as a free parameter\footnote{This approach is used because \texttt{TSFitPy} does not natively support isotope ratio fitting.}, and $A(C)$ and $A(N)$ set to the values determined in the previous step. We then fit a 4$^{\rm{th}}$ order polynomial to the $\chi^2$ vs $^{16}$O/$^{18}$O curve to determine the best-fit $^{16}$O/$^{18}$O value as the one with the minimum $\chi^2$, and $1\sigma$ uncertainties on this using the $\Delta \chi^2 = 1$ criterion (see Fig. \ref{fig:fitting}d). We then fix the $^{16}$O/$^{18}$O ratio to its best-fit value, and measure the total oxygen abundance $A(O)$ by fitting the CO absorption lines individually, similar to the measurement of C and N abundances. We report the mean and the standard-deviation of the measured line-by-line total oxygen abundances. 

    Figure \ref{fig:fitting}e shows the final fits for individual molecules.
    
\end{enumerate}

This process yields the abundances of nitrogen, oxygen, and the oxygen isotope ratio for each star in our sample, along with $1\sigma$ uncertainties for each quantity. Additionally, the atomic lines listed in Table \ref{atomiclines} are used to obtain the abundances of Fe, Mg, Na, Ca, S, and Si. {C~I lines in dLHdC stars are also analyzed as an additional estimate for the carbon abundances, discussed further in Section \ref{cprob}}. We are unable to measure accurate uncertainties for these elements because of the small number of their unblended absorption lines in our spectra. Based on the uncertainties for the CNO elements, we conservatively estimate an uncertainty of $\approx 0.2$ dex on the abundances of these elements.

\begin{table*}[]
    \centering
    \caption{Chemical abundances and oxygen isotope ratios for the stars in our sample.} \label{tab:results}
    \begin{tabular}{lcccccccccccc}
        \hline \\[-3ex]
        \hline
        \textbf{Star} & \boldmath$T_{\mathrm{eff}} \textbf{(K)}$ & \textbf{A(C)} & \textbf{A(N)} & \textbf{A(O)} & \boldmath$\mathbf{{}^{16}\mathrm{O}/{}^{18}\mathrm{O}}$ & {\textbf{A(C)*}} & \textbf{[Fe]} & \textbf{[Mg]} & \textbf{[Na]} & \textbf{[Ca]} & \textbf{[S]} & \textbf{[Si]} \\
    \hline
    HD 137613    & 5500  & 9.86 $\pm$ 0.12  & 9.24 $\pm$ 0.06   & 8.72 $\pm$ 0.08   & 0.52$_{-0.11}^{+0.14} $ & {9.6}  & -0.2  & 0.4   & 0.3   & -0.2  & -0.2   & -0.4  \\
    HD 182040    & 5750  & 9.90 $\pm$ 0.13  & 9.08 $\pm$ 0.08   & 7.99 $\pm$ 0.06   & 0.58$_{-0.24}^{+0.35} $ & {9.7}  & -0.5  & 0.2   & -0.0  & -0.4  & -0.2   & -0.2  \\
    B566         & 5750  & 10.03 $\pm$ 0.25 & 9.09 $\pm$ 0.10   & 8.55 $\pm$ 0.10   & 0.78$_{-0.23}^{+0.32} $ & {9.9}  & -0.3  & 0.0   & 0.0   & -0.5  & -0.1   & 0.3   \\
    C38          & 5750  & 10.32 $\pm$ 0.16 & 9.47 $\pm$ 0.08   & 8.99 $\pm$ 0.10   & 0.73$_{-0.21}^{+0.26} $ & {10.2} & 0.2   & 0.7   & 0.6   & 0.0   &  0.2   &  0.2   \\
    A223         & 6250  & 9.48 $\pm$ 0.24  & 8.73 $\pm$ 0.09   & 8.20 $\pm$ 0.09   & 0.60$_{-0.19}^{+0.24} $ & {9.3}  & -0.8  & -0.1  & -0.1  & -0.8  & -0.4   & -0.7  \\
    B42          & 5500  & 10.17 $\pm$ 0.14 & 9.18 $\pm$ 0.14   & 8.81 $\pm$ 0.1    & 0.32$_{-0.15}^{+0.28}$  & {9.8}  & -0.3  & 0.3   & 0.3   & 0.1   & -0.1   & 0.0   \\
    NSV11154     & 5250  & 8.90 $\pm$ 0.06  & 7.45 $\pm$ 0.15   & 7.19 $\pm$ 0.10   & 7.18$_{-2.76}^{+6.38} $ & {--}    & -0.8  & -1.2  & -1.6  & -1.5  & -0.1   & -0.8  \\
    ASAS-RCB-21  & 5000  & 8.86 $\pm$ 0.06  & 7.74 $\pm$ 0.13   & 7.77 $\pm$ 0.16   & 7.54$_{-2.53}^{+4.14} $ & {--}    & -0.9  & -1.4  & -1.2  & -1.1  & -0.6   & -1.1  \\
    ASAS-RCB-18  & 5000  & 9.15 $\pm$ 0.05  & 7.83 $\pm$ 0.19   & 8.01 $\pm$ 0.14   & 9.73$_{-3.02}^{+4.89} $ & {--}    & -0.7  & -1.2  & -1.0  & -0.9  & -0.9   & -0.4  \\
    WISE J1818+  & 5000  & 8.93 $\pm$ 0.06  & 7.60 $\pm$ 0.12   & 7.55 $\pm$ 0.10   & 7.47$_{-2.73}^{+4.27} $ & {--}    & -0.7  & -1.3  & -1.3  & -0.9  & -0.2   & -1.0  \\
    AO Her       & 4750  & 8.95 $\pm$ 0.05  & 6.95 $\pm$ 0.11   & 7.76 $\pm$ 0.14   & 69  $_{-35}^{+228}    $ & {--}    & -1.0  & -1.3  & -1.5  & -1.7  & -0.3   & -0.1  \\
    WISE J1942+  & 4500  & 8.72 $\pm$ 0.06  & 6.59 $\pm$ 0.13   & 7.42 $\pm$ 0.10   & 93  $_{-55}^{+500}    $ & {--}    & -0.9  & -1.2  & -1.8  & -1.6  &  0.0   & -0.6  \\
    \hline
    \end{tabular} \\ \vspace{1em}
    \raggedright
    \textbf{Note:} Fe, Mg, Na, Ca, S, and Si abundances are reported relative to solar values of A(Fe) = 7.5, A(Mg) = 7.6, A(Na) = 6.3, A(Ca) = 6.4, A(S) = 7.2, A(Si) = 7.6 \citep{magg2022}. {A(C)* represents the C abundance recovered from C~I lines, only measurable for the dLHdC stars.} \\
    {\textbf{Note:} In HdC stars, absolute abundances depend on the assumed C/He ratio, which cannot be directly measured from our spectra due to a lack of helium lines, potentially introducing uncertainties. The use of abundance ratios is more robust for analyzing the chemical composition of these stars (see \citealt{Asplund2000}).}
\end{table*}

The measured elemental abundances and oxygen isotope ratios for the stars in our sample are listed in Table \ref{tab:results}. The CNO abundances and oxygen isotope ratios are precisely constrained for all stars in our sample, except WISE\,J1942+. For this star, the $^{12}$C$^{18}$O lines are very weak compared to the $^{12}$C$^{16}$O lines. This causes the $\chi^2$ vs $^{16}$O/$^{18}$O curve to flatten for large oxygen isotope ratios and never rise above the $\Delta \chi^{2}=1$ level from its minimum point. As a result, while we can measure the best-fit and 1-$\sigma$ lower bound, the upper-bound on the $^{16}$O/$^{18}$O is not constrained, and we report an upper limit of 500 for this star.

We obtain consistent isotope ratio measurements for the two stars HD\,137613 and HD\,182040 that were analyzed previously using high-resolution spectra by \citet{Garcia-Hernandez2009}. The CNO abundances are slightly different because the previous study used a fixed model abundance of $A(C) = 9.5$, while we fit this using C$_2$ lines. Medium-resolution spectra of all six dLHdC stars and three RCB stars in our sample were analyzed by \citet{karambelkar2021}. The isotope ratios of three of the dLHdCs and one RCB are within the ranges reported from medium-resolution spectra, {while the remaining do not match}. The reported nitrogen abundances also have uncertainties spanning 2.5 dex in most cases. This highlights the importance of using high-resolution spectra to accurately measure abundances and isotope ratios in the \textit{K}-band due to extensive blending of molecular features.

We note that the C/N ratios of the dLHdC stars in our sample lie between 4 -- 10, while those of the RCB stars range from 13 -- 135 (the two cooler RCB stars AO Her and WISE J1942+ have ratios $>100$, while the rest are below 30). Both the carbon and nitrogen abundances are higher in the dLHdCs, but this is more prominent with nitrogen: $A(N)>8.7$ for dLHdCs and $<8.0$ for RCBs. We observe the inverse of the remark made by \citet{tisserand2022} that dLHdCs may have a lower nitrogen abundance based on a comparison of the strength of CN features observed in medium-resolution optical spectra of dLHdCs and RCBs with the same effective temperature. We propose that this discrepancy is a temperature effect, as our sample contains RCBs that are cooler than the dLHdCs. In support of this suggestion, \citet{Garcia-Hernandez2009,Garcia-Hernandez2010} reported higher nitrogen abundances for the warmer RCBs in their sample. We also note that there is no apparent distinction in the values of C/O ratios between the two classes, and these range from 13 to 80. 

We also see that all the RCBs have additional broadening of lines in their spectra, with $v_{\text{mac}}$ values ranging from 6 -- 8 km s$^{-1}$. This is not observed in the dLHdCs, and the difference is visually evident in Figures \ref{fig:CN} and \ref{fig:CO}. Our instrumental resolution of 4 km s$^{-1}$ was critical in detecting this dichotomy, and this could point at the presence of a more dynamic and turbulent atmosphere in RCBs which could be related to dust formation in these stars.

Finally, we observe that the dLHdCs seem to have a higher metallicity than the RCBs, with the mean [Fe] $= -0.3$ for the former and [Fe] $= -0.8$ for the latter. All the stars in our sample have sub-solar metallicities, with the exception of C38 which has [Fe] $= +0.2$.

\subsection{Sources of systematic errors}
We now discuss the limitations of our analysis that are likely to introduce some systematic effects in the derived elemental abundances. We note that as these limitations have similar effects on the strengths of ${}^{12}$C${}^{16}$O and ${}^{12}$C${}^{18}$O lines,  they should not significantly alter the oxygen isotope ratios which are the main focus of our work {\citep{Asplund2000,Garcia-Hernandez2009}}.

\begin{table*}[t]
    \centering
    \caption{Effect of changes in the input stellar parameters on the derived abundances of ASAS-RCB-21. {In case of $\log g$, the only other value in the model grid is 0.5 dex, and we therefore only state changes introduced by $\Delta log g = $ 0.5 dex}.}  \label{tab:paramerr}
    \begin{tabular}{lcccccc}
        \hline \\[-3ex]
        \hline
                     & Chosen Value & $\Delta$(param)       & $\Delta A(N)$ & $\Delta A(O)$ &$\Delta A(C)$ & $\Delta ^{16}\textrm{O}/^{18}\textrm{O}$ \\ 
        \hline
    $T_\textrm{eff}$ & 5000 K       & $\pm 250$ K           & $\pm$0.1     & $\pm$0.05     &  $\pm0.07$  &        $(-1.1, +1.9)$     \\
    $\log g$         & 1 dex & $0.5$ dex    & $+0.08$       & $+0.03$       &  $+0.03$  &    $-1.2$                  \\
    $v_\textrm{mic}$ & 7 km s$^{-1}$ & $\pm 2$ km s$^{-1}$   & $\pm$0.04     & $\pm$0.02     &  $\pm0.02$  &          $\pm 0.09$       \\
        \hline
    \end{tabular}
\end{table*}

\subsubsection{Model Grid}
The sparse grid of H-deficient stellar models available in the literature is a significant limitation in any elemental abundance study of HdC stars. It is clear that the range of input elemental abundances available in the model-grid are not suitable for fitting the spectra of all HdC stars in our sample. For several stars, the derived abundances differ substantially from the input values assumed in the MARCS models. A new set of HdC atmospheric models that span a wider range of elemental abundances is required to address this issue. The abundances derived in this paper will help inform the input abundance choices for new models.

The MARCS models used in our analysis were made with a fixed $A(C)$ value of 9.5 that appears to be too high for the cooler RCB stars in our sample. We find that $A(C)$ ranges from 8.7 to 10.3. Similarly, $A(O)$ is observed to range from 7.2 in the RCB NSV11154 to 9.0 in the dLHdC C38, and the mean value of 8.1 indicates that the fixed input model $A(O)$ of 8.8 is too large for most stars. The nitrogen abundances observed are well represented by the grid of models available in $A(N)$, and this work highlights the importance of generating a similar grid of models for $A(C)$ and $A(O)$. As noted earlier, most of the stars in our sample show [Fe] $<0$. It is clear that the assumed solar metallicity in the MARCS models is not appropriate for most of these stars, and an expansion of the model grid to include sub-solar metallicities is necessary.

{These discrepancies from the input model abundances are likely to introduce errors in the absolute abundances we report. While we cannot directly quantify these errors due to a lack of models, \citet{Asplund2000} tested the effects for R CrB. They generated a new model matching the values derived from a model that deviated from the abundances obtained by as much as 1.3 dex -- comparable to the differences we observe. They found that the abundances derived from the new model were within 0.05 dex of the original values, suggesting that the derived abundances are robust to deviations in the input abundances.}

\subsubsection{Dust Dilution}
\label{sec:dilution}

{RCB stars show an infrared (IR) excess due to the presence of warm circumstellar dust \citep{feast1997,Garcia-Hernandez2009,tisserand2012}. This is expected to contribute an additional continuum flux component that has a veiling effect over photospheric absorption lines, resulting in the lines appearing weaker than they actually are. This effect is evident in the spectrum of IRAS 1813+ shown in Figures \ref{fig:CN} and \ref{fig:CO}. As discussed in Section \ref{sec:observations}, this star was in a decline of $\Delta V \approx 1$ mag from its maximum luminosity, indicating that a fresh dust ejection episode from this star {has led to a diluted} photospheric spectrum, compounding the effects of the existing circumstellar dust shell. This dust shell is also expected to weaken the lines of the other RCB stars in our sample, but it is an effect that is difficult to estimate and correct for. We would require spectral energy distributions (SEDs) from the same epoch as the spectra, along with accurate models of the properties of the dust shell, which is beyond the scope of this work.}

{Instead, we attempt to quantify this effect for a few different stars using a simplified model. We examine the effect of a dust shell continuum flux that is added to the photospheric flux and estimate the true photospheric spectrum.
We define the following quantities:}
{
\begin{itemize}
    \item $F_{\text{obs}}^{\text{tot}}$: Total observed flux.
    \item $F_{\text{phot}}^{\text{tot}}$, $F_{\text{dust}}^{\text{tot}}$: Total fluxes contributed by the photosphere and dust shell, respectively.
    \item $C_{\text{phot}}$, $C_{\text{dust}}$: Continuum flux levels of the photosphere and dust shell, respectively. In the absence of any dust lines in the wavelength range of interest, we set $F_{\text{dust}}^{\text{tot}} = C_{\text{dust}}$ 
    \item $C_{\text{tot}}$: Total continuum flux level. $C_{\text{tot}} = C_{\text{phot}} + C_{\text{dust}}$
    \item $f_{\text{shell}}$: Fractional contribution of the dust shell to the total observed continuum flux (i.e. $f_\text{shell} = C_{\text{dust}} / C_{\text{tot}} = F_{\text{dust}}^{\text{tot}}/ C_{\text{tot}}$).
    \item $F_{\text{obs}}^{\text{norm}}, F_{\text{phot}}^{\text{norm}}$: Continuum-normalized observed and photospheric fluxes, respectively. These can be written as 
    $F_{\text{obs}}^{\text{norm}} = F_{\text{obs}}^{\text{tot}}/C_{\text{tot}}$; $F_{\text{phot}}^{\text{norm}} = F_{\text{phot}}^{\text{tot}}/C_{\text{phot}}$
\end{itemize}
}
{Our continuum normalized fluxes measure $F_{\text{obs}}^{\text{norm}}$ and we wish to determine $F_{\text{phot}}^{\text{norm}}$.} 

Writing the total observed flux as :
\begin{align*}
    F_{\text{obs}}^{\text{tot}} &= F_{\text{phot}}^{\text{tot}} + F_{\text{dust}}^{\text{tot}} \\
                                &= F_{\text{phot}}^{\text{tot}} + f_\text{shell} \cdot C_{\text{tot}}
\end{align*}

Rearranging this gives :

\begin{equation*}
    F_{\text{phot}}^{\text{tot}} = F_{\text{obs}}^{\text{tot}} - f_\text{shell} \cdot C_{\text{tot}}
\end{equation*}

Finally, dividing by C$_{\text{phot}}= (1-f_\text{shell}) \cdot C_{\text{tot}}$ to convert to a continuum-normalized flux and simplifying gives the final relation:

\begin{equation}
    \boxed{
    F_{\text{phot}}^{\text{norm}} = \frac{ F_{\text{obs}}^{\text{norm}} - f_\text{shell} }{ 1 - f_\text{shell} }
    }
\end{equation}

{Given a value of $f_{\text{shell}}$, we can now determine the true, undiluted continuum-normalized photospheric flux.
It is important to note that $f_{\text{shell}}$, like the individual continuum levels, is a wavelength-dependent quantity that we assume to be fixed for simplicity.}
We chose $f_\textrm{shell} = 20\%$ as a reasonable value to test, as we estimate this quantity to be in the 10 -- 30$\%$ range for these stars {in the \textit{K}-band. We detect no trend of the recovered abundances with wavelength, indicating that the gradient in $f_{\text{shell}}$ is much smaller than the line-to-line scatter, and that fixing $f_{\text{shell}}$ is a reasonable approximation.} After applying this correction we observed an increase in depth of all lines, with stronger lines being affected the most. For most of the species that we analyze, the lines appear to be far from saturation, and we see a change of less than $0.1$ dex for all the elemental abundances other than that of oxygen. In the case of oxygen, the \coa lines are approaching saturation, and the abundance of $^{16}$O is therefore affected strongly, with the correction causing an increase of up to 0.4 dex depending on the star. However, since the \cob lines are rather weak, the change in abundance of $^{18}$O is similarly small like the other elements. This leads to an increase of 0.45 -- 0.55 dex in $A(O)$ which is dominated by the $^{16}$O fraction increasing, and an accompanying increase in the $^{16}$O/$^{18}$O ratio by roughly a factor of 2. This indicates that the circumstellar dust shell has a significant effect on estimates of the oxygen isotope ratio, and is a factor that must be studied carefully in future works that use near-IR spectra of RCB stars.

\subsubsection{Uncertainties in Stellar Parameters}
\label{sec:errors}

Table \ref{tab:paramerr} lists the errors introduced by small changes in the input parameters of the models. We checked the effects of changing T$_\textrm{eff}$, $\log g$, and $v_{\text{mic}}$ on the derived abundances for the star ASAS-RCB-21. There appears to be no detectable change for small variations of $v_{\text{mac}}$.
Based on these results, we estimate systemic errors of 0.08, 0.13, and, 0.06 dex for the carbon, nitrogen, and oxygen abundances respectively, and $ \pm 1.5$ on the oxygen isotope ratio for this star. These are consistent with the uncertainties reported in Table \ref{tab:results}. 

\subsubsection{Carbon Problem} \label{cprob}

One major issue encountered while studying HdC stars is the so-called `carbon problem' that was extensively discussed by \citet{Asplund2000} in their detailed analyses of high-resolution optical spectra of warm (T$_\textrm{eff}$ > 6250 K) RCB stars. The issue arises because the photoionization of neutral carbon atoms, that are in {highly excited states} in such warm atmospheres, is the dominant source of continuum opacity. This leads to a discrepancy between the predicted and the observed C~I lines strength. The carbon abundances derived from C~I absorption lines are systematically lower by a median value of $\sim$0.6 dex than the input abundances in the stellar atmospheric models \citep{Asplund2000}.

To get around the `carbon problem', \citet{Hema2012,Hema2017} used the C$_2$ lines to measure the carbon abundance from their optical high resolution spectra. This technique appeared to be an excellent alternative. However the estimates with C$_2$ lines resulted in values systematically lower than the ones found with C I lines, implying that the ratio C/He in HdC stars' atmosphere should be much lower that the ratio of $\sim$0.6\% measured in {Extreme Helium} (EHe) stars \citep{Asplund2000,Pandey2006}. Consequently, they noted that, if one considers that both groups of stars should have similar modes of formation, a version of the carbon problem with the C$_2$ lines could also affect their measurements. Overall, using warm and cold HdC stars, they estimated that the carbon abundance should be lower for all HdC stars, and extrapolated low C/He ratios ranging between 0.03\% and 0.3\%. Furthermore, it is worth noting that, with their colder HdC stars sample, their measure of the carbon abundance from the C I lines and from the C$_2$ lines agreed (their Table 6), indicating possibly a variation of the `carbon problem' with effective temperature.

\begin{figure*}[t]
    \centering
    \includegraphics[width=0.57\textwidth]{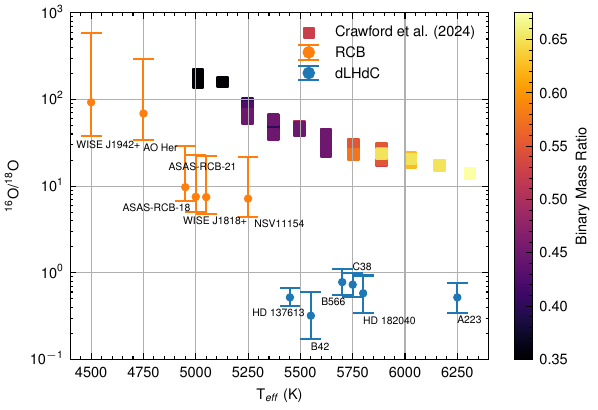}
    \includegraphics[width=0.405\textwidth]{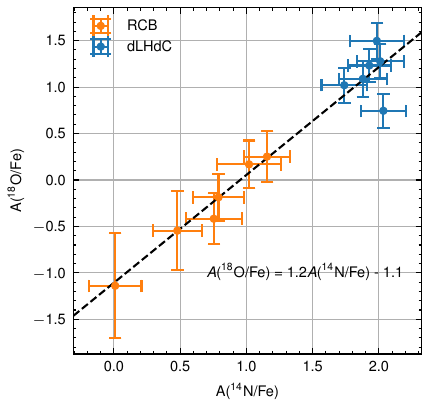}
        \caption{\emph{Left:} Variation of oxygen isotope ratio with $T_{\mathrm{eff}}$ for the RCB stars (orange dots) and dLHdC stars (blue dots) in our sample. In cases where multiple stars have the same $T_{\mathrm{eff}}$, offsets of 50 K have been applied for readability. We find that all RCB stars in our sample have $^{16}$O/$^{18}$O $>4$, while dLHdC stars have $^{16}$O/$^{18}$O $<1$. We also find a decreasing trend of the oxygen isotope ratios with effective temperature, consistent with predictions from WD merger remnant models from \citet{Crawford2024} (colored squares, color-coded by the mass-ratio of the progenitor WD binary). {Error bars on oxygen isotope ratios of RCBs are increased to include the possible effects of dust shell dilution (see Sec \ref{sec:dilution})} \\ 
        \emph{Right:} Abundances of ${}^{18}$O with N (right) {relative to Fe} for the dLHdC (blue) and RCB (orange) stars in our sample. 
        We find that {$A({}^{18}\textrm{O}/\textrm{Fe})$ ($=A({}^{18}\textrm{O}$) - A(\textrm{Fe}))} is linearly correlated with {$A({}^{14}\textrm{N}/\textrm{Fe})$} (black dashed line), suggesting that a fixed fraction {($\approx 8\%$)} of the nitrogen is converted to $^{18}$O across all stars via $\alpha$-capture.
        } 
        \label{fig:isorat}
\end{figure*}

\citet{Garcia-Hernandez2010} underlined that, depending on the HdC star effective temperature, the main contributor to the continuum opacity may change to free-free absorption from neutral helium when observing in the infrared K band, thus the `carbon problem' could disappear at such wavelengths. \citet{Garcia-Hernandez2009, Garcia-Hernandez2010} measured the carbon abundance for 5 cold RCB, 2 cold dLHdC and 2 warm dLHdC (T$_\textrm{eff}$ > 6250 K) stars from high resolution infrared spectra. 
They reported not seeing the `carbon problem' with any of the HdC stars using the C I lines, and the measured carbon abundances were consistent with the input model (C/He=1\%), within a range of -0.3 to +0.4 dex. However, they observed a `carbon problem' with the cold HdC stars using the C$_2$ lines. Their measured carbon abundance was about 0.8 dex higher than the input abundance.

Similar to \citet{Garcia-Hernandez2009, Garcia-Hernandez2010}, we observed cold HdC stars in the infrared. We used the new C$_2$ ExoMol linelist \citep{2018MNRAS.480.3397Y} to estimate the carbon abundances from C$_2$ lines. For RCB stars, our derived C abundances range between 8.7 and 9.2, which is $\sim$0.6 dex lower than the input abundance of 9.52 (C/He=1\%). 
{As noted in Section \ref{sec:dilution}, correcting for dilution by circumstellar dust could increase $A(C)$ by about 0.1 dex, which would reduce this gap slightly.}
However, conversely, for dLHdCs, our measured C abundances exceed the input abundance by 0.4 to 0.8 dex, which is similar to the result reported by \citet{Garcia-Hernandez2009}. For the two dLHdC stars we have in common, HD 137613 and HD 182040, we derived a C$_2$ based carbon abundance that are about 0.5 dex lower than the ones reported by \citet{Garcia-Hernandez2009}. That is possibly explained by the use of different C$_2$ line lists.

The difference in C/He ratio we are observing between the RCB stars and the dLHdC stars may be related to a difference in the formation scenario. Both of them are expected to originate from WD mergers, but accumulating evidences show that RCB and dLHdC stars could be formed by different WD systems in term of total mass and mass ratios \citep{tisserand2022, Crawford2024}.

{Lastly, we are also able to estimate the carbon abundance from C~I lines (A(C)* in Table \ref{tab:results}) that are visible in the spectra of all the dLHdC stars in our sample but missing from the RCBs (likely due to a combination of lower carbon abundances and their cooler temperatures favoring C$_2$ formation). We found these to be systematically lower than the values obtained using C$_2$ lines, with the difference ranging from -0.42 to -0.13 with an average value of -0.23. We note that the uncertainties on these differences are up to 0.2 -- 0.3 dex, so all the stars appear to be consistent with each other. This appears to disagree with the results reported by \citet{Hema2012}, who observed the C$_2$ abundances to be lower than C~I for warm RCB stars using optical spectra. This discrepancy could again be related to a variation of the `carbon problem' with effective temperature and/or wavelength.}

An expanded set of models with a grid of input carbon abundances is required to quantify the extent of the carbon problem with C$_2$ lines{, and verify the hypothesis that the differences in the C/He ratios of RCB and dLHdC stars could be related to their formation scenarios}. As the main focus of our study is oxygen isotope ratios, we leave this analysis for a future study and report the carbon abundances derived with our current models in Table 4.

\section{Discussion}

\label{sec:discussion}

The main goal of this paper was to measure precise oxygen isotope ratios for a large sample of RCB and dLHdC stars. Using our high-resolution spectra, we find that all dLHdC stars in our sample have oxygen isotope ratios lower by an order of magnitude than all RCB stars in our sample. All dLHdC stars in our sample have ${}^{16}$O/${}^{18}$O~$<1$, while all RCB stars have ${}^{16}$O/${}^{18}$O~$>4$. The coolest star in our sample RCB WISE\,J1942+ has a ratio as high as 92. {As discussed in Section \ref{sec:dilution}, ${}^{16}$O/${}^{18}$O~estimates may be underestimated by a factor of 2 for RCBs if the circumstellar dust shell contributes to 20\% of the observed \textit{K}-band flux. This would further widen the ${}^{16}$O/${}^{18}$O gap between RCB and dLHdC stars.}
Our results robustly confirm that dLHdC stars tend to have lower ${}^{16}$O/${}^{18}$O ratios than RCB stars -- a trend suggested by previous analyses of medium resolution spectra \citep{karambelkar2022} and small samples of high-resolution spectra \citep{Garcia-Hernandez2010}. As noted previously, this difference could potentially be explained by requiring that dLHdCs and RCBs originate in white-dwarf binaries with distinct properties such as total masses or mass-ratios (see \citealt{tisserand2022,karambelkar2022, Crawford2023} for details). 

Additionally, we note a correlation between the oxygen isotope ratios and the effective temperatures of HdC stars (Figure \ref{fig:isorat} left), with warmer stars having lower $^{16}$O/$^{18}$O ratios than the cooler ones. The two coolest stars ($T_{\mathrm{eff}} \sim 4500$ K) have oxygen-isotope ratios that are an order of magnitude larger than the warmer ($T_{\mathrm{eff}} > 5000$ K) stars. This is the first observed trend of oxygen isotope ratios in RCB and dLHdC stars. This trend of oxygen isotope ratios with the effective temperatures of HdC stars has also been predicted by the recent \citet{Crawford2024} theoretical models. They find that the lower $q$ ($= \textrm{M}_\textrm{donor}/\textrm{M}_\textrm{accretor}$), higher M$_{\rm{tot}}$ mergers generally produce colder stars with higher oxygen isotope ratios, while high $q$, low M$_{\rm{tot}}$ mergers produce hotter stars with low oxygen isotope ratios. Fig. \ref{fig:isorat} (left) also shows the oxygen isotope ratios as a function of effective temperature from the \citet{Crawford2024} model grid, highlighting a dependence on $q$. It is promising that the models predict the same general trend seen in the observations. However, as noted in \citet{Crawford2024}, the exact values of the oxygen isotope ratios are not reproduced by the models. We note that the models overpredict the observed oxygen isotope ratios of RCBs by a factor of $\approx2.5$ {(this could potentially be bridged by the effects of dilution, as noted in Section \ref{sec:dilution})} and dLHdCs by two orders of magnitude.  \citet{Crawford2024} suggest that the model oxygen isotope ratios can be reduced by adopting lower helium-shell burning temperatures (T$_{\rm{SOF}}$), but the lowest ratio achieved for dLHdC models was $3.76$, which is still significantly larger than our observations. Additionally the amount of hydrogen in the He-WD or the extent of convection can also impact the oxygen isotope ratios \citep{Munson2021, Zhang14}. Further modeling to reproduce the observed values of the isotope ratios will shed light on the role of the WD progenitor properties (in addition to M$_{\rm{tot}}$ and $q$) in determining the observed properties of RCB and dLHdC stars. 

We also note a positive correlation between the abundances of ${}^{14}$N and ${}^{18}$O in the HdC stars (Figure \ref{fig:isorat}, right). This is consistent with the $\alpha$-capture reaction chain \hbox{$^{14}$N$(\alpha, \gamma)$$^{18}$F$(\beta^{+}\nu)^{18}$O} believed to be responsible for the formation of $^{18}$O during partial He burning \citep{clayton2005}. A linear fit to the data in Figure \ref{fig:isorat} (right) gives {a slope of $1.16 \pm 0.13$ and a y-intercept of $-1.11 \pm 0.22$ dex}, suggesting that the amount of $^{18}$O produced is proportional to the amount of $^{14}$N present, and that {$\approx 8\%$} of $^{14}$N is converted to $^{18}$O across all HdC stars. {Dust shell dilution does not play a significant role here as the increase in these abundances was found to be $\sim 0.1$ dex, which is further mitigated by our use of abundance ratios.}

\section{Conclusion}
\label{sec:conclusion}
In this paper, we have analyzed high-resolution (R$\approx75000$) K-band (2.25 -- 2.48 \um) spectra of six RCB and six dLHdC stars to determine their chemical abundances and oxygen isotope ratios. We fit the observed spectra to synthetic spectra constructed using a grid of MARCS models for HdC stars and precisely measured their ${}^{16}$O/${}^{18}$O ratios. We find that all six RCB stars in our sample have ${}^{16}$O/${}^{18}$O $>4$ and all six dLHdC stars have ${}^{16}$O/${}^{18}$O $<1$, confirming previous suggestions that dLHdC stars have lower oxygen isotope ratios than RCB stars. 
{In our investigation of the possible effects of dilution caused by circumstellar dust in RCBs, we showed that the oxygen isotope ratios of RCB stars could be underpredicted by a factor of 2 if the dust emission contributes to 20\% of the observed \textit{K}-band flux, highlighting the need for careful characterization of this dust in future studies of RCBs that rely on infrared spectra.}
Additionally, for the first time, we find a possible correlation between the oxygen isotope ratios and the effective temperatures of HdC stars, with colder stars having larger $^{16}$O/$^{18}$O values than their hotter counterparts. This trend is consistent with recent simulations of white-dwarf mergers that explore the effects of total mass and mass ratios of WD binaries on their merger remnants. However, these models overpredict the $^{16}$O/$^{18}$O values by two orders of magnitude, pointing to additional factors that could play an important role in setting the oxygen isotope ratios.

We also used our high-resolution spectra to measure abundances of C, N, O, Fe, S, Si, Mg, Na, and Ca for the twelve HdC stars. We find a linear relation between the ${}^{18}$O abundance and the ${}^{14}$N abundance, suggesting that a fixed fraction ($\approx 8 \%$) of ${}^{14}$N is converted by $\alpha$-capture to ${}^{18}$O in all HdC stars. We also find that all RCB stars in our sample have a lower metallicity than the dLHdC stars. 

Our work highlights the need for new, publicly available MARCS models with a wider grid of input abundances to accurately model the spectra of HdC stars. An extension to other $A(C)$, $A(O)$, and [Fe] values is necessary to successfully analyze this diverse group of stars. Additionally, all RCB stars in our sample are cooler than 5250~K, while the dLHdC stars are warmer than 5500~K, and this has a potential to bias our results. Future high resolution observations of a more diverse sample of stars will verify the trends reported in this study. 

In summary, we are arriving at a consistent picture for HdC stars, where the observed differences between RCBs and dLHdCs such as their effective temperatures, luminosities, and oxygen isotope ratios can be attributed to differences in the properties of their progenitor white-dwarf binaries.

\section{Acknowledgements}
AM and PT thank Sergey Yurchenko from the ExoMol team for providing diatomic molecular line lists, and Bertrand Plez for the discussions we had during the course of this article. PT thanks the MARCS team in Uppsala (Sweden) and especially Kjell Eriksson for kindly providing a grid of hydrogen-deficient stellar models. AM thanks Nick Storm and Jeffrey Gerber for suggestions that led to the method we developed to fit isotope ratios. MMK and VB thank the Department of Science and Technology, Government of India, and Indian National Academy of Engineering for an inaugural Vaishvik Bharatiya Vaigyanik (VAIBHAV) fellowship. {This work has made use of the VALD database, operated at Uppsala University, the Institute of Astronomy RAS in Moscow, and the University of Vienna. The observations presented in this paper were obtained under the Visiting Astronomer Program at the Infrared Telescope Facility, which is operated by the University of Hawaii under contract 80HQTR24DA010 with the National Aeronautics and Space Administration. This publication makes use of data products from the Near-Earth Object Wide-field Infrared Survey Explorer (NEOWISE), which is a joint project of the Jet Propulsion Laboratory/California Institute of Technology and the University of California, Los Angeles. NEOWISE is funded by the National Aeronautics and Space Administration. The authors wish to recognize and acknowledge the very significant cultural role and reverence that the summit of Maunakea has always had within the Native Hawaiian community. We are most fortunate to have the opportunity to conduct observations from this mountain.} \\

\section{Software and Data availability}
The high-resolution spectra of all HdC stars and the software used for abundance and isotope ratio measurements in this paper are available at this github repository : \href{https://github.com/advaitmehla/HdC-high-res}{\texttt{https://github.com/advaitmehla/HdC-high-res}}.

\bibliography{myreferences}

\begin{thebibliography}{}
\expandafter\ifx\csname natexlab\endcsname\relax\def\natexlab#1{#1}\fi

\bibitem[{{Alvarez} \& {Plez}(1998)}]{ts0}
{Alvarez}, R., \& {Plez}, B. 1998, \aap, 330, 1109

\bibitem[{{Asplund} {et~al.}(2000){Asplund}, {Gustafsson}, {Lambert}, \& {Rao}}]{Asplund2000}
{Asplund}, M., {Gustafsson}, B., {Lambert}, D.~L., \& {Rao}, N.~K. 2000, \aap, 353, 287

\bibitem[{{Bell} {et~al.}(1976){Bell}, {Eriksson}, {Gustafsson}, \& {Nordlund}}]{bell1976}
{Bell}, R.~A., {Eriksson}, K., {Gustafsson}, B., \& {Nordlund}, A. 1976, \aaps, 23, 37

\bibitem[{{Bergeat} {et~al.}(2001){Bergeat}, {Knapik}, \& {Rutily}}]{bergeat2001}
{Bergeat}, J., {Knapik}, A., \& {Rutily}, B. 2001, \aap, 369, 178

\bibitem[{{Clayton}(1996)}]{clayton1996}
{Clayton}, G.~C. 1996, \pasp, 108, 225

\bibitem[{{Clayton}(2012)}]{2012JAVSO..40..539C}
---. 2012, \jaavso, 40, 539

\bibitem[{{Clayton} {et~al.}(2007){Clayton}, {Geballe}, {Herwig}, {Fryer}, \& {Asplund}}]{clayton2007}
{Clayton}, G.~C., {Geballe}, T.~R., {Herwig}, F., {Fryer}, C., \& {Asplund}, M. 2007, \apj, 662, 1220

\bibitem[{{Clayton} {et~al.}(2005){Clayton}, {Herwig}, {Geballe}, {Asplund}, {Tenenbaum}, {Engelbracht}, \& {Gordon}}]{clayton2005}
{Clayton}, G.~C., {Herwig}, F., {Geballe}, T.~R., {et~al.} 2005, \apjl, 623, L141

\bibitem[{{Crawford} {et~al.}(2024){Crawford}, {Nikultsev}, {Clayton}, {Tisserand}, {Soon}, \& {Pedersen}}]{Crawford2024}
{Crawford}, C.~L., {Nikultsev}, N., {Clayton}, G.~C., {et~al.} 2024, \mnras, 534, 1018

\bibitem[{{Crawford} {et~al.}(2023){Crawford}, {Tisserand}, {Clayton}, {Soon}, {Bessell}, {Wood}, {Garc{\'\i}a-Hern{\'a}ndez}, {Ruiter}, \& {Seitenzahl}}]{Crawford2023}
{Crawford}, C.~L., {Tisserand}, P., {Clayton}, G.~C., {et~al.} 2023, \mnras, 521, 1674

\bibitem[{{Cushing} {et~al.}(2004){Cushing}, {Vacca}, \& {Rayner}}]{spextool}
{Cushing}, M.~C., {Vacca}, W.~D., \& {Rayner}, J.~T. 2004, \pasp, 116, 362

\bibitem[{{Feast} {et~al.}(1997){Feast}, {Carter}, {Roberts}, {Marang}, \& {Catchpole}}]{feast1997}
{Feast}, M.~W., {Carter}, B.~S., {Roberts}, G., {Marang}, F., \& {Catchpole}, R.~M. 1997, \mnras, 285, 317

\bibitem[{{Garc{\'\i}a-Hern{\'a}ndez} {et~al.}(2009){Garc{\'\i}a-Hern{\'a}ndez}, {Hinkle}, {Lambert}, \& {Eriksson}}]{Garcia-Hernandez2009}
{Garc{\'\i}a-Hern{\'a}ndez}, D.~A., {Hinkle}, K.~H., {Lambert}, D.~L., \& {Eriksson}, K. 2009, \apj, 696, 1733

\bibitem[{{Garc{\'\i}a-Hern{\'a}ndez} {et~al.}(2010){Garc{\'\i}a-Hern{\'a}ndez}, {Lambert}, {Kameswara Rao}, {Hinkle}, \& {Eriksson}}]{Garcia-Hernandez2010}
{Garc{\'\i}a-Hern{\'a}ndez}, D.~A., {Lambert}, D.~L., {Kameswara Rao}, N., {Hinkle}, K.~H., \& {Eriksson}, K. 2010, \apj, 714, 144

\bibitem[{{Garc{\'\i}a-Hern{\'a}ndez} {et~al.}(2023){Garc{\'\i}a-Hern{\'a}ndez}, {Rao}, {Lambert}, {Eriksson}, {Reddy}, \& {Masseron}}]{agh2023}
{Garc{\'\i}a-Hern{\'a}ndez}, D.~A., {Rao}, N.~K., {Lambert}, D.~L., {et~al.} 2023, \apj, 948, 15

\bibitem[{{Gerber} {et~al.}(2023){Gerber}, {Magg}, {Plez}, {Bergemann}, {Heiter}, {Olander}, \& {Hoppe}}]{ts20}
{Gerber}, J.~M., {Magg}, E., {Plez}, B., {et~al.} 2023, \aap, 669, A43

\bibitem[{{Goorvitch}(1994)}]{1994ApJS...95..535G}
{Goorvitch}, D. 1994, \apjs, 95, 535

\bibitem[{{Gustafsson} {et~al.}(1975){Gustafsson}, {Bell}, {Eriksson}, \& {Nordlund}}]{gust1975}
{Gustafsson}, B., {Bell}, R.~A., {Eriksson}, K., \& {Nordlund}, A. 1975, \aap, 42, 407

\bibitem[{{Gustafsson} {et~al.}(2008){Gustafsson}, {Edvardsson}, {Eriksson}, {J{\o}rgensen}, {Nordlund}, \& {Plez}}]{gust2008}
{Gustafsson}, B., {Edvardsson}, B., {Eriksson}, K., {et~al.} 2008, \aap, 486, 951

\bibitem[{{Hema} {et~al.}(2017){Hema}, {Pandey}, {Kamath}, {Kameswara Rao}, {Lambert}, \& {Woolf}}]{Hema2017}
{Hema}, B.~P., {Pandey}, G., {Kamath}, D., {et~al.} 2017, \pasp, 129, 104202

\bibitem[{{Hema} {et~al.}(2012){Hema}, {Pandey}, \& {Lambert}}]{Hema2012}
{Hema}, B.~P., {Pandey}, G., \& {Lambert}, D.~L. 2012, \apj, 747, 102

\bibitem[{{Jeffery} {et~al.}(2011){Jeffery}, {Karakas}, \& {Saio}}]{Jeffery2011}
{Jeffery}, C.~S., {Karakas}, A.~I., \& {Saio}, H. 2011, \mnras, 414, 3599

\bibitem[{{Kameswara Rao} {et~al.}(2004){Kameswara Rao}, {Reddy}, \& {Lambert}}]{rao2004}
{Kameswara Rao}, N., {Reddy}, B.~E., \& {Lambert}, D.~L. 2004, \mnras, 355, 855

\bibitem[{{Kameswara Rao} {et~al.}(1999){Kameswara Rao}, {Lambert}, {Adams}, {Doss}, {Gonzalez}, {Hatzes}, {James}, {Johns-Krull}, {Luck}, {Pandey}, {Reinsch}, {Tomkin}, \& {Woolf}}]{rao1999}
{Kameswara Rao}, N., {Lambert}, D.~L., {Adams}, M.~T., {et~al.} 1999, \mnras, 310, 717

\bibitem[{{Karambelkar} {et~al.}(2022){Karambelkar}, {Kasliwal}, {Tisserand}, {Clayton}, {Crawford}, {Anand}, {Geballe}, \& {Montiel}}]{karambelkar2022}
{Karambelkar}, V., {Kasliwal}, M.~M., {Tisserand}, P., {et~al.} 2022, \aap, 667, A84

\bibitem[{{Karambelkar} {et~al.}(2021){Karambelkar}, {Kasliwal}, {Tisserand}, {De}, {Anand}, {Ashley}, {Delacroix}, {Hankins}, {Jencson}, {Lau}, {McKenna}, {Moore}, {Ofek}, {Smith}, {Soria}, {Soon}, {Tinyanont}, {Travouillon}, \& {Yao}}]{karambelkar2021}
{Karambelkar}, V.~R., {Kasliwal}, M.~M., {Tisserand}, P., {et~al.} 2021, \apj, 910, 132

\bibitem[{{Karambelkar} {et~al.}(2024){Karambelkar}, {Kasliwal}, {Tisserand}, {Anand}, {Ashley}, {Bildsten}, {Clayton}, {Crawford}, {De}, {Earley}, {Hankins}, {Hall}, {Lamberts}, {Lau}, {McKenna}, {Moore}, {Ofek}, {Smith}, {Soria}, {Soon}, \& {Travouillon}}]{Karambelkar2024}
---. 2024, \pasp, 136, 084201

\bibitem[{{Kurucz}(2010)}]{K10}
{Kurucz}, R.~L. 2010, Robert L. Kurucz on-line database of observed and predicted atomic transitions

\bibitem[{{Lambert} \& {Rao}(1994)}]{Lambert1994}
{Lambert}, D.~L., \& {Rao}, N.~K. 1994, Journal of Astrophysics and Astronomy, 15, 47

\bibitem[{{Magain}(1984)}]{1984A&A...134..189M}
{Magain}, P. 1984, \aap, 134, 189

\bibitem[{{Magg} {et~al.}(2022){Magg}, {Bergemann}, {Serenelli}, {Bautista}, {Plez}, {Heiter}, {Gerber}, {Ludwig}, {Basu}, {Ferguson}, {Gallego}, {Gamrath}, {Palmeri}, \& {Quinet}}]{magg2022}
{Magg}, E., {Bergemann}, M., {Serenelli}, A., {et~al.} 2022, \aap, 661, A140

\bibitem[{{Mainzer} {et~al.}(2014){Mainzer}, {Bauer}, {Cutri}, {Grav}, {Masiero}, {Beck}, {Clarkson}, {Conrow}, {Dailey}, {Eisenhardt}, {Fabinsky}, {Fajardo-Acosta}, {Fowler}, {Gelino}, {Grillmair}, {Heinrichsen}, {Kendall}, {Kirkpatrick}, {Liu}, {Masci}, {McCallon}, {Nugent}, {Papin}, {Rice}, {Royer}, {Ryan}, {Sevilla}, {Sonnett}, {Stevenson}, {Thompson}, {Wheelock}, {Wiemer}, {Wittman}, {Wright}, \& {Yan}}]{mainzer2014}
{Mainzer}, A., {Bauer}, J., {Cutri}, R.~M., {et~al.} 2014, \apj, 792, 30

\bibitem[{{Munson} {et~al.}(2021){Munson}, {Chatzopoulos}, {Frank}, {Clayton}, {Crawford}, {Denissenkov}, \& {Herwig}}]{Munson2021}
{Munson}, B., {Chatzopoulos}, E., {Frank}, J., {et~al.} 2021, \apj, 911, 103

\bibitem[{{Pandey} {et~al.}(2006){Pandey}, {Lambert}, {Jeffery}, \& {Rao}}]{Pandey2006}
{Pandey}, G., {Lambert}, D.~L., {Jeffery}, C.~S., \& {Rao}, N.~K. 2006, \apj, 638, 454

\bibitem[{{Plez}(2008)}]{plez2008}
{Plez}, B. 2008, Physica Scripta Volume T, 133, 014003

\bibitem[{{Plez}(2012)}]{ts1}
---. 2012, {Turbospectrum: Code for spectral synthesis}, Astrophysics Source Code Library, record ascl:1205.004

\bibitem[{{Rayner} {et~al.}(2022){Rayner}, {Tokunaga}, {Jaffe}, {Bond}, {Bonnet}, {Ching}, {Connelley}, {Cushing}, {Kokubun}, {Lockhart}, {Vacca}, \& {Warmbier}}]{ishell}
{Rayner}, J., {Tokunaga}, A., {Jaffe}, D., {et~al.} 2022, \pasp, 134, 015002

\bibitem[{{Ryabchikova} {et~al.}(2015){Ryabchikova}, {Piskunov}, {Kurucz}, {Stempels}, {Heiter}, {Pakhomov}, \& {Barklem}}]{vald}
{Ryabchikova}, T., {Piskunov}, N., {Kurucz}, R.~L., {et~al.} 2015, \physscr, 90, 054005

\bibitem[{{Storm} \& {Bergemann}(2023)}]{tsfit}
{Storm}, N., \& {Bergemann}, M. 2023, \mnras, 525, 3718

\bibitem[{Tennyson {et~al.}(2024)Tennyson, Yurchenko, Zhang, Bowesman, Brady, Buldyreva, Chubb, Gamache, Gorman, Guest, Hill, Kefala, Lynas-Gray, Mellor, McKemmish, Mitev, Mizus, Owens, Peng, Perri, Pezzella, Polyansky, Qu, Semenov, Smola, Solokov, Somogyi, Upadhyay, Wright, \& Zobov}]{TENNYSON2024109083}
Tennyson, J., Yurchenko, S.~N., Zhang, J., {et~al.} 2024, Journal of Quantitative Spectroscopy and Radiative Transfer, 326, 109083

\bibitem[{{Tisserand}(2012)}]{tisserand2012}
{Tisserand}, P. 2012, \aap, 539, A51

\bibitem[{{Tisserand} {et~al.}(2024){Tisserand}, {Crawford}, {Soon}, {Clayton}, {Ruiter}, \& {Seitenzahl}}]{tisserand2024}
{Tisserand}, P., {Crawford}, C.~L., {Soon}, J., {et~al.} 2024, \aap, 684, A131

\bibitem[{{Tisserand} {et~al.}(2020){Tisserand}, {Clayton}, {Bessell}, {Welch}, {Kamath}, {Wood}, {Wils}, {Wyrzykowski}, {Mr{\'o}z}, \& {Udalski}}]{tiss2020}
{Tisserand}, P., {Clayton}, G.~C., {Bessell}, M.~S., {et~al.} 2020, \aap, 635, A14

\bibitem[{{Tisserand} {et~al.}(2022){Tisserand}, {Crawford}, {Clayton}, {Ruiter}, {Karambelkar}, {Bessell}, {Seitenzahl}, {Kasliwal}, {Soon}, \& {Travouillon}}]{tisserand2022}
{Tisserand}, P., {Crawford}, C.~L., {Clayton}, G.~C., {et~al.} 2022, \aap, 667, A83

\bibitem[{{Vacca} {et~al.}(2003){Vacca}, {Cushing}, \& {Rayner}}]{xtellcor}
{Vacca}, W.~D., {Cushing}, M.~C., \& {Rayner}, J.~T. 2003, \pasp, 115, 389

\bibitem[{{Yurchenko} {et~al.}(2018){Yurchenko}, {Szab{\'o}}, {Pyatenko}, \& {Tennyson}}]{2018MNRAS.480.3397Y}
{Yurchenko}, S.~N., {Szab{\'o}}, I., {Pyatenko}, E., \& {Tennyson}, J. 2018, \mnras, 480, 3397

\bibitem[{{Zhang} {et~al.}(2014){Zhang}, {Wang}, {Mazzali}, {Bai}, {Zhang}, {Bersier}, {Huang}, {Fan}, {Mo}, {Wang}, {Yi}, {Wang}, {Xin}, {Liangchang}, {Zhang}, {Lun}, {Wang}, {He}, \& {Walker}}]{Zhang14}
{Zhang}, J., {Wang}, X., {Mazzali}, P.~A., {et~al.} 2014, \apj, 797, 5

\end{thebibliography}
\bibliographystyle{apj}

\section{Appendix}

\begin{table}[h]
    \begin{tabular}{lccccc}
    \hline \\[-3ex]
    \hline
    Wavelength ($\angstrom$) & Lower Energy Level (eV) & $\log gf$ & Excitation Potential (eV) & $\log \gamma$ & Broadening (s$^{-1}$) \\
    \hline
    22906.458          & 9.1718                  & -0.279    & 2.5                       & 0.0           & $3.98\times 10^8$              \\
    23062.949          & 10.3851                 & -1.842    & 2.5                       & 3.0           & $3.98\times 10^8$              \\
    23091.400          & 10.3524                 & -1.494    & 2.5                       & 2.0           & $3.98\times 10^8$              \\
    23091.827          & 10.3524                 & -0.987    & 2.5                       & 2.0           & $3.98\times 10^8$              \\
    23154.750          & 9.1718                  & -1.736    & 2.5                       & 0.0           & $3.98\times 10^8$              \\
    23178.263          & 10.5202                 & -2.385    & 2.5                       & 1.0           & $3.98\times 10^8$              \\
    23178.371          & 10.4081                 & -1.124    & 2.5                       & 3.0           & $3.98\times 10^8$              \\
    23438.000            & 10.352                  & -0.748    & 2.5                       & 0.0           & $3.98\times 10^8$              \\
    23442.268          & 10.3524                 & -1.569    & 2.5                       & 2.0           & $3.98\times 10^8$              \\
    23442.818          & 10.3524                 & -0.659    & 2.5                       & 0.0           & $3.98\times 10^8$              \\
    23580.101          & 10.1979                 & -1.585    & 2.5                       & 0.0           & $3.98\times 10^8$              \\
    23607.814          & 10.5306                 & -2.081    & 2.5                       & 1.0           & $3.98\times 10^8$              \\
    24442.939          & 9.6311                  & -0.851    & 2.5                       & 2.0           & $3.98\times 10^8$              \\
    24981.651          & 10.3851                 & -0.182    & 2.5                       & 3.0           & $3.98\times 10^8$             \\
    \hline
    \end{tabular}
    \caption{The C~I linelist used by us, modified from the VALD-Kurucz \citep{vald,K10} list to remove lines that were missing from our data.}
    \label{tab:clist}
\end{table}

\label{lastpage}
\end{document}